\documentclass[prb,twocolumn,showpacs,superscriptaddress]{revtex4}

\usepackage{graphicx}
\usepackage{amsmath}
\usepackage{amssymb}
\usepackage{epsfig}

\renewcommand{\v}[1]{{\bf #1}}

\def\eqa{\begin{eqnarray}}
\def\eea{\end{eqnarray}}
\newcommand{\eq}{\begin{equation}}
\newcommand{\ee}{\end{equation}}
\newcommand{\nn}{\nonumber\\}
\newcommand{\Eq}[1]{Eq.~(\ref{#1})}
\newcommand{\<}{\langle}
\renewcommand{\>}{\rangle}

\newcommand{\p}{\partial}

\newcommand{\ua}{\uparrow}
\newcommand{\da}{\downarrow}
\newcommand{\ra}{\rightarrow}

\newcommand{\al}{\alpha}
\newcommand{\bt}{\beta}

\newcommand{\Del}{\Delta}

\newcommand{\ga}{\gamma}
\newcommand{\Ga}{\Gamma}

\newcommand{\La}{\Lambda}

\newcommand{\si}{\sigma}

\usepackage{color}

\begin{document}

\title{Superconductivity in Sr$_2$RuO$_4$ thin film under biaxial strain}

\author{Yuan-Chun Liu}
\affiliation{National Laboratory of Solid State Microstructures $\&$ School of Physics, Nanjing
University, Nanjing, 210093, China}

\author{Wan-Sheng Wang}
\affiliation{Department of Physics, Ningbo University, Ningbo 315211, China}

\author{Fu-Chun Zhang}
\affiliation{Kavli Institute for Theoretical Sciences, University of Chinese Academy of Sciences, Beijing 100190, China}
\affiliation{Collaborative Innovation Center of Advanced Microstructures, Nanjing 210093, China}

\author{Qiang-Hua Wang}
\affiliation{National Laboratory of Solid State Microstructures $\&$ School of Physics, Nanjing
University, Nanjing, 210093, China}
\affiliation{Collaborative Innovation Center of Advanced Microstructures, Nanjing 210093, China}
\email{qhwang@nju.edu.cn}


\begin{abstract}

Motivated by the success of experimental manipulation of the band structure through biaxial strain in Sr$_2$RuO$_4$ thin film grown on a mismatched substrate, we investigate theoretically the effects of biaxial strain on the electronic instabilities, such as superconductivity (SC) and spin density wave (SDW), by functional renormalization group. According to the experiment, the positive strain (from lattice expansion) causes charge transfer to the $\ga$-band and consequently Lifshitz reconstruction of the Fermi surface. Our theoretical calculations show that within a limited range of positive strain a $p$-wave superconducting order is realized. However, as the strain is increased further the system develops into the SDW state well before the Lifshitz transition is reached. We also consider the effect of negative strains (from lattice constriction). As the strain increases, there is a transition from $p$-wave SC state to nodal $s$-wave SC state. The theoretical results are discussed in comparison to experiment and can be checked by further experiments.

\end{abstract}

\pacs{71.10.Fd, 74.20.-z, 74.20.Rp, 71.27.+a}
%

\maketitle

\section{Introduction}

Sr$_2$RuO$_4$ is a leading candidate that possibly hosts $p+ip'$-wave superconductivity.\cite{Mackenzie, Bergemann, Maeno2012, Kallin2012}. In agreement with the initial theoretical proposals ~\cite{RiceSigrist,Baskaran}, various experiments provide evidence for odd parity Cooper pairs~\cite{liu}, with total spin equal to one~\cite{NMR}, and  chiral time-reversal breaking symmetry~\cite{muSR,Kerr}. The chiral $p+ip'$-wave superconductor is of great current interest because of its topological property that may lead to zero energy Majorana bound states~\cite{ivanov}, the building block for topological quantum computing~\cite{nayak}.

Since the superconducting transition temperature ($T_c\sim 1.5$K) is fairly low in the parent Sr$_2$RuO$_4$, it is of interest to achieve higher $T_c$ by tuning the filling and/or bandstructure of the material.
The Fermi surface of Sr$_2$RuO$_4$ has two distinct components \cite{Mackenzie,Bergemann}, two approximately 1-dimensional (1D) $\alpha$ and $\beta$ bands and a single 2D $\gamma$ band. Out of these bands, the $\ga$ band has a van Hove singularity slightly above the Fermi level, and hence is more relevant to superconductivity and more viable to tuning of the band structure. Doping toward the van Hove singularity would cause higher density of states (DOS) at the Fermi level and would enhance $T_c$. However, doping by direct chemical substitution is not promising since Sr$_2$RuO$_4$ is extremely sensitive to disorder that would be caused by dopants. \cite{Mackenzie1998} Instead, strain can be used to tune the material without causing disorders. Recently, Hicks {\em et al} \cite{Hicks} and Steppke {\em et al} \cite{Steppke} applied both compressive and tensile uniaxial strain to find an increase of ${T_c}$ up to twice of that in unstrained sample, followed by a sudden drop as the strain is increased further. An interesting issue is whether the $T_c$-peak is associated with the Lifshitz transition in the $\ga$ band. The resistivity data appears to be consistent with the scenario that the $T_c$-peak corresponds to the Lifshitz transition within superconducting state, while no other electronic ordering occurs.\cite{transport} Since the unstrained sample has spin triplet pairing, and only the spin singlet pairing is favored at the Lifshitz point, this scenario indicates a transition from spin triplet to spin singlet states at the Lifshitz point \cite{Steppke}, which is interesting, but would need further evidence from other types of experiments.  In other scenario, the $T_c$ peak is the result of phase transition from superconductivity to spin-density-wave order.\cite{YCL}  The latter is ferromagnetic-like, which would only lead to majority and minority bands without opening quasiparticle gap. We may argue that such a spin-density-wave may not lead to a drastic change in the resistivity, hence may not be inconsistent with the resistivity data in Ref.\cite{transport} In principle, a van Hove singularity at the Fermi level could be probed more effectively by using spectroscopy measurement, although it may be challenging for the uniaxially strained samples.

On the other hand, it is now possible to grow thin films of Sr$_2$RuO$_4$ on various mismatched substrates, \cite{Tokura,Burganov} leading to biaxial strain from elongated Ru-O bond. The biaxial strain is seen to cause charge transfer from the $\al$ and $\bt$ bands to the $\ga$ band up to and beyond the Lifshitz transition.\cite{Burganov} Unfortunately, superconductivity is not yet observed in the biaxially strained samples. While further refinement of the thin film quality may be necessary, it is pertinent to investigate the effect of the biaxial strain theoretically to find the optimal level of biaxial strain, given the sudden drop of $T_c$ in the case of uniaxial strain.\cite{Steppke} In Ref.\cite{eaKim} the effect of biaxial strain is discussed within a three-band model by applying a theoretical scheme that is exact in the limit of infinitesimal interaction. However, the possibility of competing orders is not addressed, which is important at finite interactions that are known to lead to significant renormalization of the effective mass in Sr$_2$RuO$_4$. Here we apply the singular-mode functional renormalization group (SM-FRG) \cite{WSW1, WSW2, WSW3, YYX, YangYang, YCL, WWW}, which is an unbiased method capable of dealing with competing orders on equal footing. We notice that there are still concerns on whether the 2D $\gamma$ band or the 1D $\alpha$ and $\beta$ bands are active for superconductivity,\cite{Huo2013, Raghu2010, Chung2012}, but the previous FRG study by including all of the three bands show that the $\ga$ band plays a dominant role in the superconducting state \cite{YangYang}. Combined with the sensitivity of the $\ga$ band versus the strain as revealed by the experiments, we will limit ourselves to the $\ga$ band for simplicity.

The main results of this paper are as follows. Within a limited range of positive strain (from lattice expansion) a $p$-wave superconducting (SC) order is realized, with increasing transition temperature due to increasing DOS at the Fermi level. However, the system enters the spin-density-wave (SDW) state well before the Lifshitz transition is reached.
However, if the system is in a parameter space where singlet SC is realized in the unstrained case, the system may keep singlet SC as the biaxial strain increases and even pushes the system to the van Hove singularity. While this case can not be ruled out on a pure theoretical basis, it has to reconcile with previous evidences of $p$-wave pairing.
We also consider the effect of negative strain (from lattice constriction). The transition temperature of the $p$-wave SC state decreases, and as the strain magnitude increases further the system enters a nodal $s$-wave SC state.

The rest of this paper is arranged as follows. In Sec.\ref{M&M} we describe the model and the effect of biaxial strain on the band structure. We also describe SM-FRG briefly, leaving the technical details in the Appendix. In Sec.\ref{R&D} we describe the results from SM-FRG. Finally we summarize and discuss the results in Sec.\ref{SMR}.

\section{Model and Method} \label{M&M}

In the experiment, \cite{Burganov} Sr$_2$RuO$_4$ thin films grown on mismatched substrates has enlarged lattice constants. We assume that under such a positive biaxial strain, all hopping integrals are reduced by a common factor. If on the other hand the fillings on the three orbitals were unchanged, the free part of the hamiltonian would be the same as that for the parent compound upon a rescaling of hoppings and chemical potential. This is however not the case in experiment. In fact, by calculating the Luttinger volume from the angle-resolved photo-emission spectroscopy it turns out that while the total filling is barely changed, there is a charge transfer from the $d_{xz}$/$d_{yz}$ orbitals to the $d_{xy}$ orbital, leading to increasing filling of the $\gamma$ band as the (positive) strain increases (versus different substrates). Therefore, a reasonable model for the $\gamma$ band, which we argued previously as the most relevant one for superconductivity, is  described by the following hamiltonian,
\eqa
H = &&-\sum_{\< ij \> \si}(c^\dag_{i\si}t_{ij}c_{j\si}+{\rm h.c.})-\mu\sum_{i\si}n_{i\si}\nn
&&+U\sum_i n_{i\ua}n_{i\da}+V\sum_{\< ij \>\in {\rm NN}}n_i n_j. \label{H}
\eea
Here $c^\dag_{i\sigma}$/$c_{i\sigma}$ creates/annihilates an electron with spin $\si$ at site $i$, and $\<ij\>$ denotes the nearest-neighbor (NN) and the next-nearest-neighbor (NNN) bonds, with the corresponding hopping integrals $t_1$ and $t_2$. By rescaling the energy, we use dimensionless units so that $t_1=0.8$ and $t_2=0.35$, as used for the parent compound, \cite{YCL} with the understanding that they decrease uniformly with positive biaxial strain in absolute units. The chemical potential $\mu$ is to be tuned to increase the filling of the $\ga$ band, reflecting the effect of positive biaxial strain, which may be obtained by comparison to the corresponding experimental filling level. We set $\mu=1.3$ for the unstrained case in the parent compound by matching the Fermi surface of the $\gamma$ band in the parent compound Sr$_2$RuO$_4$.\cite{Mackenzie,Bergemann}

The onsite Hubbard interaction $U$ and the NN Coulomb interaction $V$ may also change under the strain, but we leave them as parameters. The interactions in the quantum many-body system can lead to instabilities of the normal state toward various electron orders. In order to treat all possible and competing electronic orders on equal footing, we apply the singular-mode functional renormalization group (SM-FRG). \cite{WSW1, WSW2, WSW3, YYX, YangYang, YCL, WWW}  Here we outline the necessary ingredients and notations, leaving technical details in the Appendix. In a nutshell, the idea is to obtain momentum-resolved running pseudo-potential $\Ga_{1234}$, as in $(1/2)c_{1\si}^\dagger c_{2\si'}^\dagger \Ga_{1234} c_{3\si'} c_{4\si}$, to act on low-energy fermionic degrees of freedom up to a cutoff energy scale $\La$ (for Matsubara frequency in our case). Here the numerical index labels momentum/position (but will be suppressed wherever applicable for brevity). Momentum conservation/translation symmetry is also left implicit. Starting from $\Ga$ at $\La\ra \infty$ (specified by the bare interactions $U$ and $V$), FRG generates all one-particle-irreducible corrections to $\Ga$ to arbitrary orders in the bare interactions as $\La$ decreases. Notice that $\Ga$ may evolve to be nonlocal and even diverging. To see the instability (diverging) channel, we extract from $\Ga$ the effective interactions versus the running scale $\La$ in the general CDW/SDW/SC channels,
 \eqa
 && V^{\rm CDW}_{(14)(32)} = 2 \Ga_{1234} - \Ga_{1243},\nn &&  V^{\rm SDW}_{(13)(42)} = -V^{\rm SC}_{(12)(43)}=- \Ga_{1234}. \label{eq:VX}
 \eea
Here $V^{\rm CDW/SDW/SC}$  are understood as matrices with composite indices, describing scattering of fermion bilinears. Since they all originate from $\Ga$, they have overlaps but are naturally treated on equal footing. Since the collective momentum of the bilinear is conserved (in a given scattering channel), the decomposition is performed at each collective momentum separately.  We remark that the FRG flow would be equivalent to ladder or random-phase approximations in the respective channels if the overlaps were ignored in the FRG flow equation. The divergence of the leading attractive (i.e., negative) eigenvalue $S$ of $V^{\rm CDW/SDW/SC}$ decides the instability channel, the associated eigenfunction and collective momentum describe the order parameter, and the divergence energy scale $\La_c$ is representative of the transition temperature $T_c$.
More technical details can be found in Refs.\cite{WSW1, WSW2, WSW3, YYX, YangYang, YCL, WWW} and also in the self-complete Appendix.

\section{Results and Discussions} \label{R&D}

We now present the results obtained by SM-FRG. We begin with case studies at several typical filling levels, mimicking different levels of biaxial strain in the sense described above, with fixed $ (U,V)=(3.5,0.7)$, and we end up with phase diagrams and discussions on the robustness of the results versus the interaction parameters.

\begin{figure}
	\includegraphics[width=8.5cm,trim={5.6cm 2.6cm 5.5cm 1.3cm},clip]{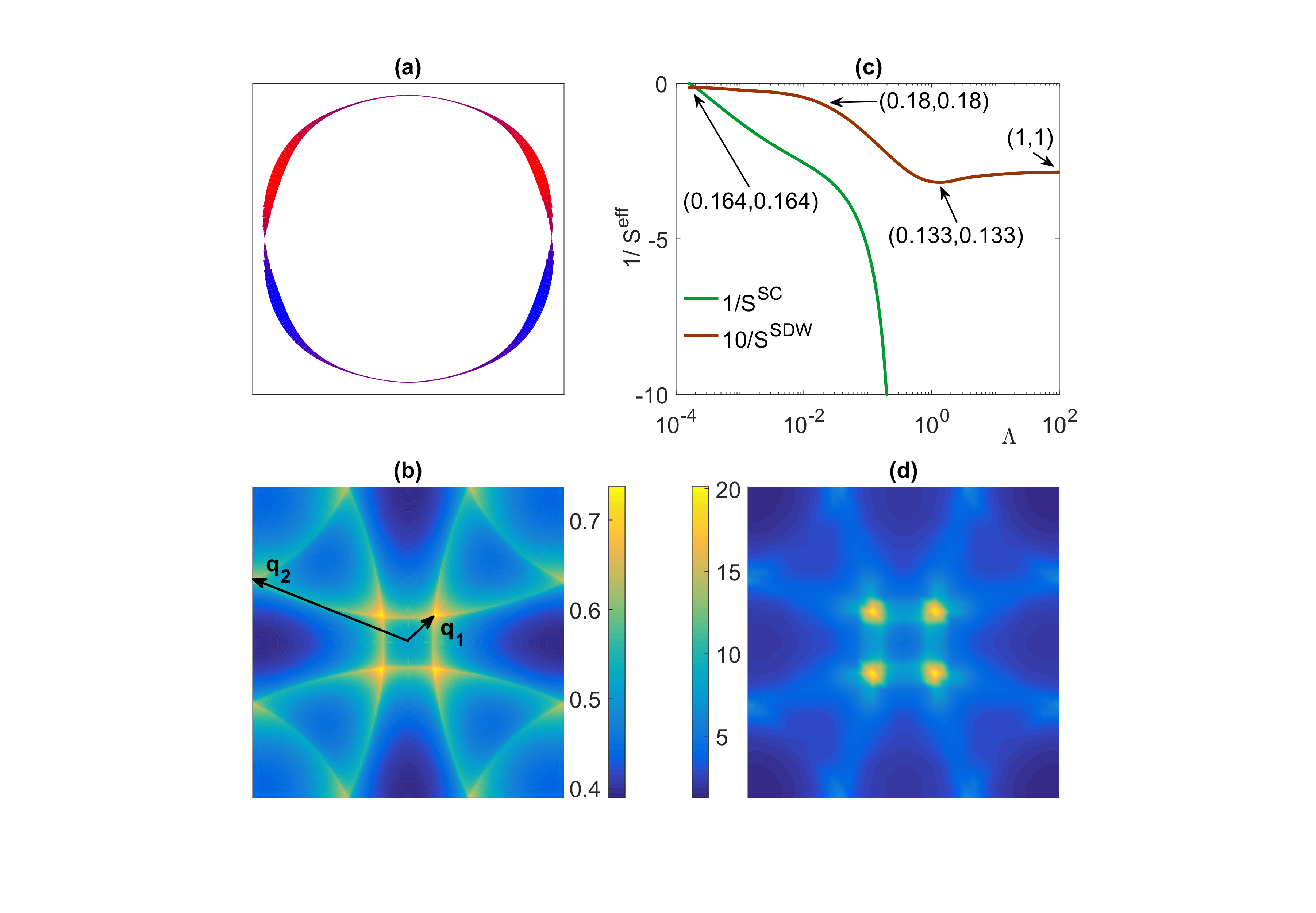}
	\caption{ (Color online) Results for $\mu = 1.31$, very weak positive strain case. (a) The Fermi surface. The gap function is also shown on the Fermi surface, where the width represents the amplitude, and the color indicates the sign (red/blue for positive/negative). (b) Bare susceptibility $\chi_0(\v q)$ versus the wavevector $\v q$. The arrows indicate strong peaks at a small (large) wavevector  $\v q_1$ ($\v q_2$).  (c) FRG flow in the SC and SDW channels versus decreasing $\La$. The arrows snapshot the momentum $\v Q$ (divided by $\pi$) associated with the leading SDW eigenmode. (d) $-V_{\rm SDW}(\v q)$ at the late stage of FRG flow.}\label{mu131}
\end{figure}

{\em Weak positive strain}: Since the unstrained case $\mu=1.3$ has been studied previously,\cite{YCL} here we consider the case of weak positive strain by setting $\mu = 1.31$. This applies to the case in which the lattice constant in the substrate is only slightly larger than that of Sr$_2$RuO$_4$. Since the van Hove level is already close to the Fermi level in the unstrained case, the system is sensitive to a small change of $\mu$. The Fermi surface is shown in Fig.\ref{mu131}(a), which is closer to the van Hove points on the zone boundary than the unstrained case, leading to larger density of states at the Fermi level. Fig.\ref{mu131}(b) shows the bare susceptibility as a function of momentum. The cuts of strong intensity trace basically $2\v k_f$-scattering, truncated in the reduced Brillouine zone. Here $\v k_f$ is the Fermi momentum. The strength is relatively stronger at the crossing points $\v q_1$ and $\v q_2$ where the scattering phase space is larger. The peak at $\v q_1$ is the strongest as a result of particle-hole scattering near the same van Hove point.

As interactions are switched on, we show in Fig.\ref{mu131}(c) the FRG flow of leading eigenvalues of the running interactions in the SC and SDW channels (plot as $1/S^{\rm X}$ for X = SC, SDW) defined in the previous section. The CDW channel remains weak during the flow and is henceforth ignored whereever applicable. $S^{\rm SDW}$ is initially stronger at high energy scales, is enhanced in the intermediate stage, and saturates eventually at low energy scales. The arrows in Fig.\ref{mu131}(c) are snapshots of collective momentum $\v Q$ (divided by $\pi$) associated with the leading SDW eigenmode, showing saturation at $\v q\sim \v q_1$.

As the SDW interaction is enhanced, it triggers attractive $S^{\rm SC}$ in the pairing channel (as a manifestation of channel overlap), which eventually diverges at a low energy scale. The pairing function is given by the eigenfunction of $V^{SC}$ with the most negative eigenvalue $S^{\rm SC}$. We find one such function is $\Del_\v k \sim \sin k_y (1-a\cos k_x)$ up to a global scale, with $a\sim 0.9$, arising from anti-phase pairing on NN and NNN bonds. (The pairing on longer bonds is negligibly small.) This is a $p_y$-wave pairing gap, and is plot on the Fermi surface in Fig.\ref{mu131}(a) for illustration. Notice that while the node along $k_y=0$ is dictated by the $p_y$-wave symmetry, the function also diminishes along $k_x=0$ because of the $a$-term in the gap function, forming a quasi-node or deep minimum in the gap amplitude. If $a=1$ this gap function would behave as $k_x^2 k_y$ in the limit of $k\ra 0$. The quasi-node along $k_x=0$ is a manifestation of the destructive effect from the proximity to the van Hove points on the zone boundary: equal-spin triplet pairing right at the van Hove momenta $\pm \v k_v$ is forbidden by Pauli principle since $\pm \v k_v$ are identical up to a umklapp vector.
The other degenerate pairing function behaves as $\Del_\v k\sim \sin k_x (1-a\cos k_y)$ (not shown). The degeneracy is guaranteed by the $C_{4v}$ point group. In the ordered state below the transition temperature, the time-reversal breaking $p_x\pm ip_y$ pairing is energetically more favorable, and in our case the gap function would behave as $k_x k_y (k_x \pm i k_y)$ if $a=1$. This is exactly the so-called $f$-wave-like gap that would be most consistent with the recent thermal conductivity experiment. \cite{kappa} However, no symmetry requires $a=1$, although in our case $a$ is close to unity. The nodal or quasi-nodal property of the gap function also agrees with earlier measurements showing abundance of low energy quasiparticle excitations deep in the superconducting state.\cite{Nishizaki1, Nishizaki2, Deguchi1, Deguchi2}

Fig.\ref{mu131}(d) shows $S^{\rm SDW}$ as a function of momentum $\v q$ at the final stage of FRG. We see that the SDW interaction at $\v q_1$ is enhanced most significantly. Together with the enhancement in the intermediate energy scales, this implies that the small-$\v q$ SDW fluctuations are compatible with or beneficial for the $p$-wave triplet pairing.
\begin{figure}
	\includegraphics[width=8.5cm,trim={5.6cm 2.6cm 5.5cm 1.3cm},clip]{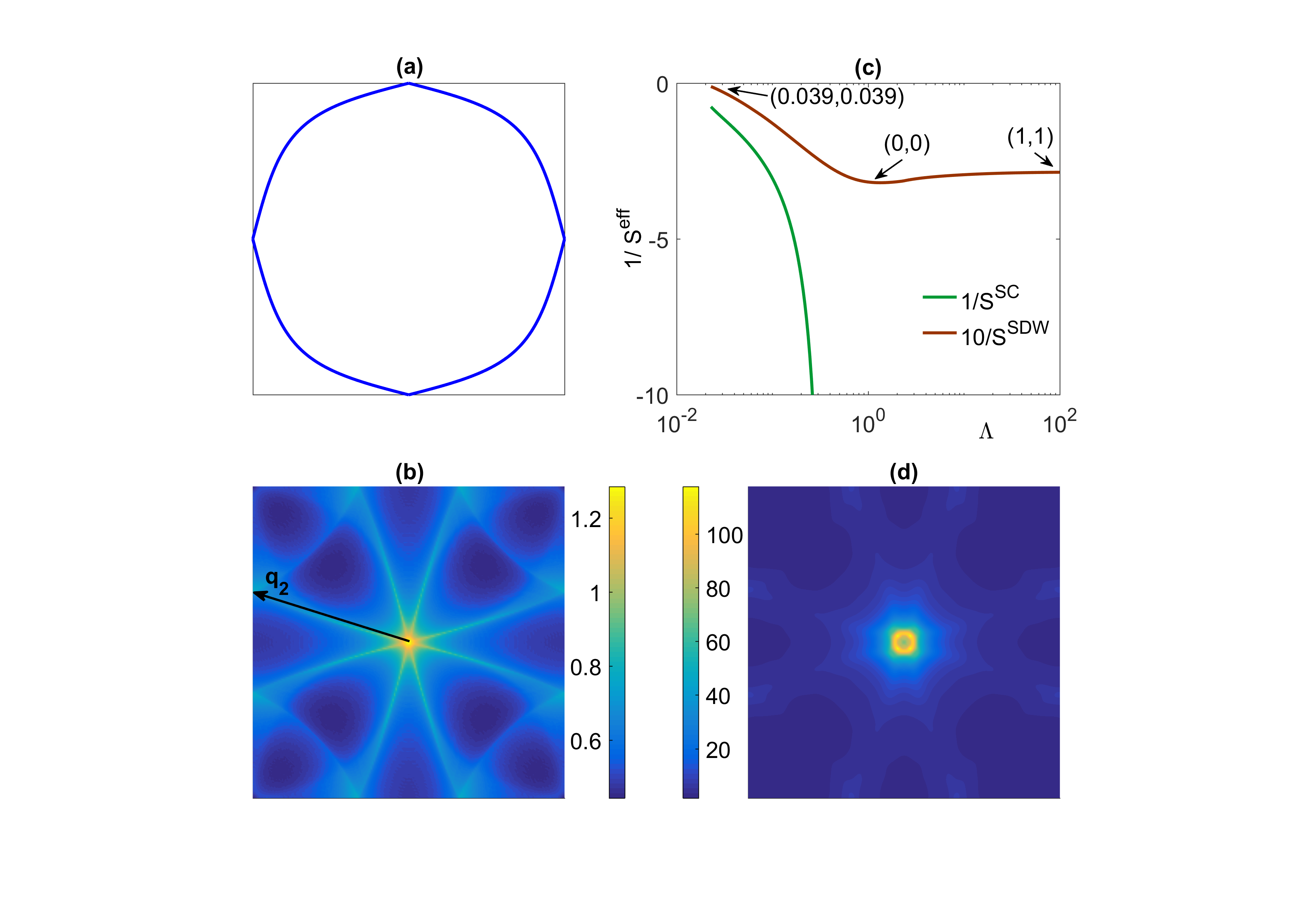}
	\caption{ (Color online) The results for $\mu = 1.4$, the case of positive strain up to the van Hove level. (a) The Fermi surface. (b) Bare susceptibility $\chi_0(\v q)$ versus the wavevector $\v q$. The arrows
		indicate strong peaks at a large wavevector  $\v q_2$. (c) FRG flow versus decreasing $\La$.  (d) $-V_{\rm SDW}(\v q)$ at the late stage of flow.} \label{mu140}
\end{figure}

\begin{figure}
	\includegraphics[width=8.5cm,trim={5.6cm 2.6cm 5.5cm 1.3cm},clip]{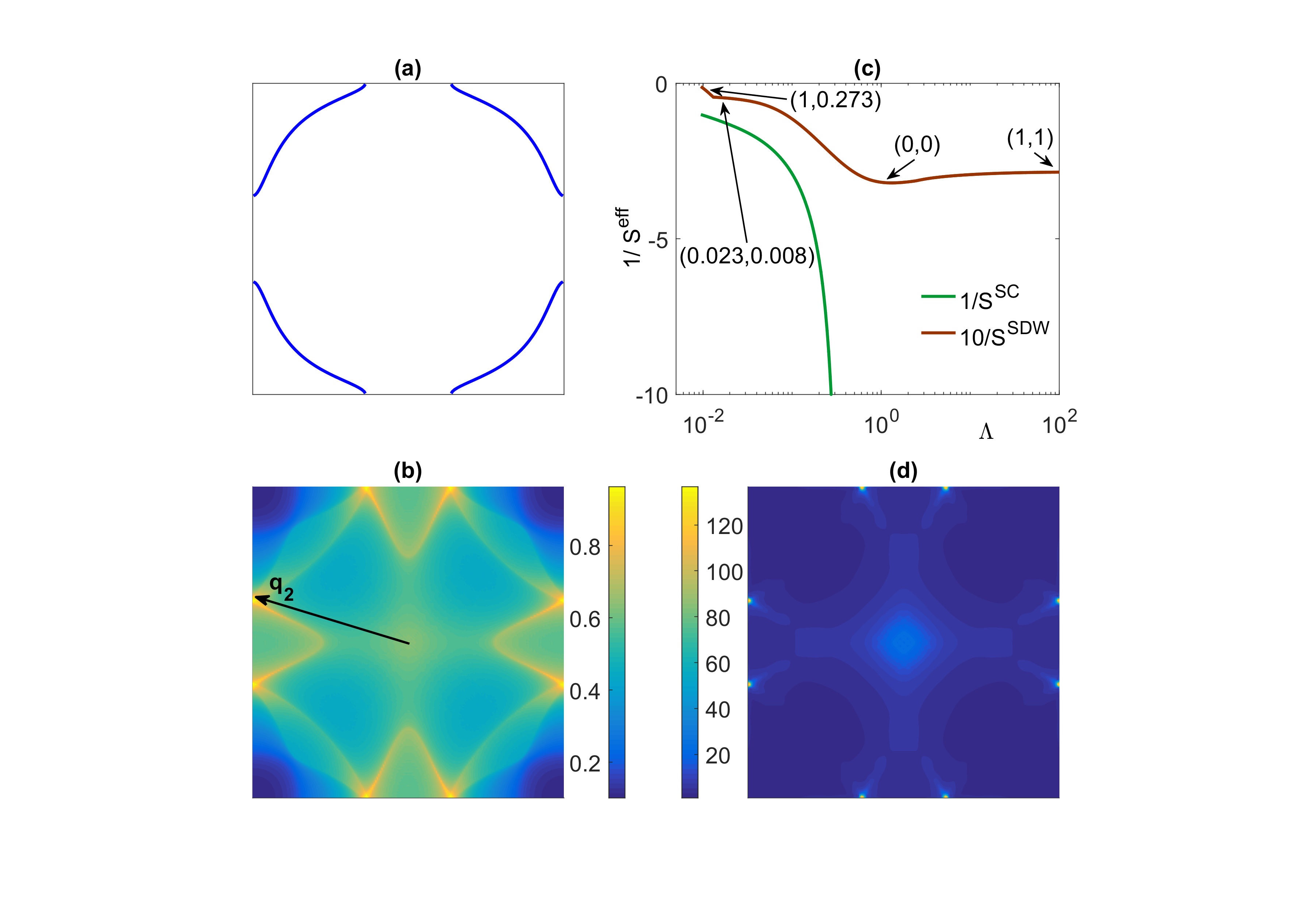}
	\caption{ (Color online) Results for $\mu = 1.47$, with positive strain well above the van Hove level. (a) Fermi surface. (b) Bare susceptibility $\chi_0(\v q)$ versus the wavevector $\v q$. The arrows
		indicate strong peaks at a large wavevector  $\v q_1$. (c) FRG flow versus
		decreasing $\La$. (d) $-V_{\rm SDW}(\v q)$ at the late stage of flow.} \label{mu147}
\end{figure}

{\em Strain up to the van Hove level}: Next we consider a biaxial strain that pushes the Fermi level up to the van Hove singularity. This corresponds to $\mu = 4t_2=1.4$. The Fermi surface is shown in Fig. \ref{mu140}(a). The bare susceptibility in Fig.\ref{mu140}(b) shows a strong peak at the origin $\v q_1=0$, aside from weaker local peak at $\v q_2$. The overall feature can again be traced to $2\v k_f$-scattering and the enhancement at $\v q_1$ is a result of van Hove singularity. The effective interactions flow as shown in Fig. \ref{mu140}(c). The SDW channel is initially stronger, is enhanced in the intermediate stage, and finally diverges, although the SC channel is triggered attractive and grows below the intermediate energy scales. Fig. \ref{mu140} (d) shows that in the last stage the SDW interaction is peaked around $\v q_1=0$. (The small deviation to the origin follows from the fact that the van Hove singularity is cut off by the fairly high divergence scale.)  We checked the decomposition of $V^{\rm SDW}$ to find that the leading eigenfunction corresponds to site-local spin density. Therefore, for the moderately strong bare interaction under concern, the system develops ferromagnetic SDW order at low temperatures. The fact that SC is less favorable here is again due to the destructive effect from the van Hove points.

{\em Strain well above the van Hove level}: If the strain is sufficiently large, it is possible to push the Fermi level above the van Hove level. Here we consider $\mu = 1.47$. The Fermi surface breaks into four arcs, as shown in Fig. \ref{mu147}(a). The bare susceptibility is now strongest at $\v q_2$, as shown in Fig.\ref{mu147}(b).
The FRG flow is shown in Fig. \ref{mu147}(c). The SDW channel behaves similarly to the previous cases at high energy scales, where the flow is insensitive to the Fermi surface. However, there is a level-crossing (see the snapshots of $\v q$) at low energy scales, where quasiparticle excitations are sensitive to the Fermi surface topology, to SDW interaction at $\v q_2$. The SC channel is subleading, although it is triggered attractive in the intermediate energy window. The momentum dependence of the SDW interaction is shown in Fig.\ref{mu147}(d), which peaks at $\v q_2$. Therefore, the system now develops SDW at a large momentum $\v q_2$.

\begin{figure}
	\includegraphics[width=8.5cm,trim={5.6cm 2.6cm 5.5cm 1.3cm},clip]{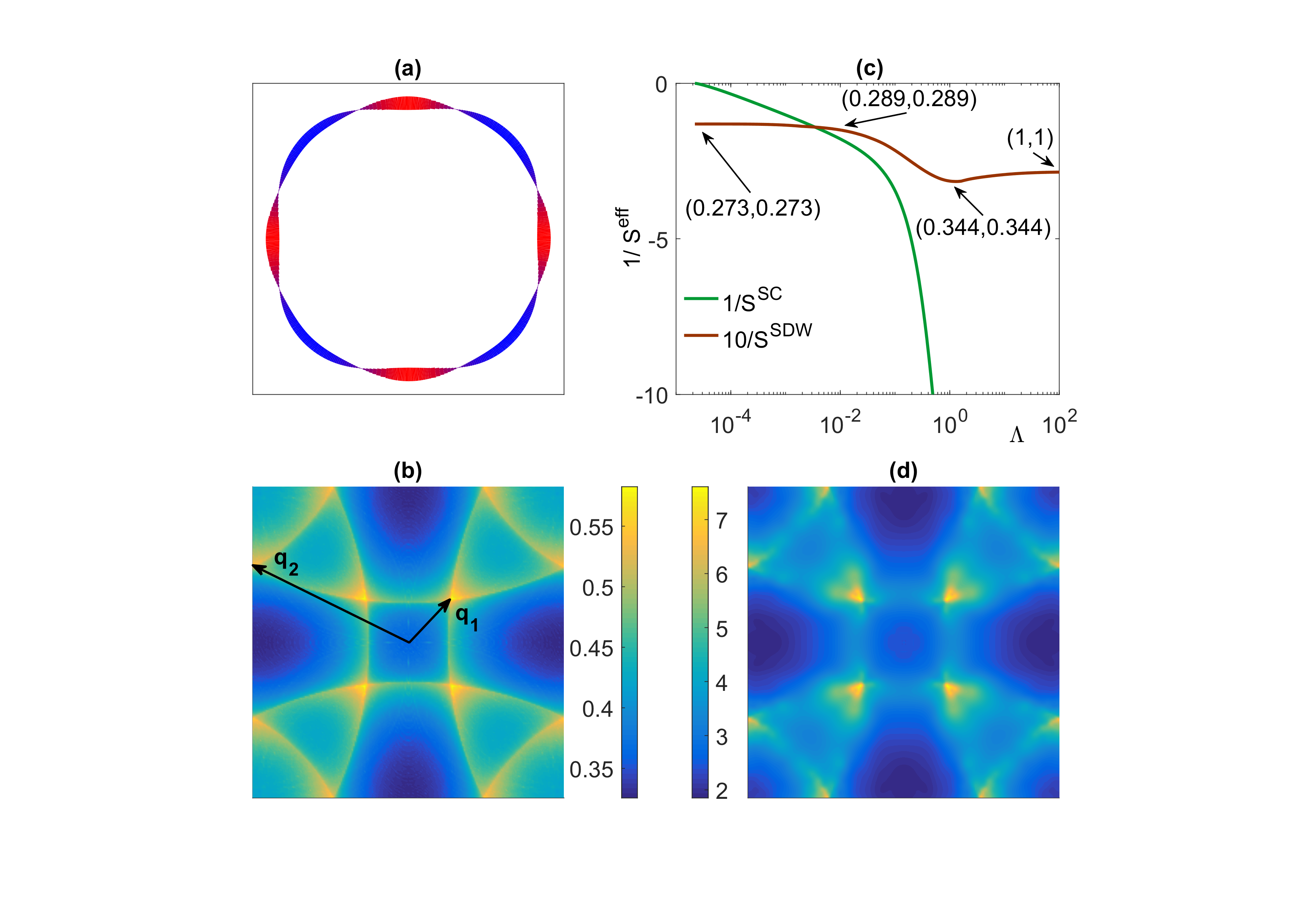}
	\caption{ (Color online) Results for $\mu = 1.16$, with strong negative strain. (a) The Fermi surface. The gap function is also shown, where the width represents the amplitude and the
     color indicates the sign (red/blue for positive/negative). (b) Bare susceptibility $\chi_0(\v q)$ versus the wavevector $\v q$. The arrows indicate strong
     peaks at a small (large) wavevector  $\v q_1$ ($\v q_2$).(c) FRG flow versus
		decreasing $\La$. (d) $-V_{\rm SDW}(\v q)$
		at the late stage of flow.} \label{mu116}
\end{figure}

{\em Negative strain}: For completeness, we also consider a negative strain that would appear from a smaller substrate lattice constant. We assume that this would cause charge transfer from the $\ga$ band to the $\al$ and $\bt$ bands. To mimic this situation, we consider $\mu = 1.16$, well below the chemical potential (for the $\ga$ band) in the unstrained case. The Fermi surface is shown in Fig.\ref{mu116}(a). The bare susceptibility in Fig. \ref{mu116}(b) shows peaks at $\v q_1$ and $\v q_2$. Here $\v q_1$ is larger than that in the previous cases. The FRG flow is shown in Fig.\ref{mu116}(c).  As in previous cases, the SDW channel triggered attractive pairing interactions in the intermediate energy window. Eventually, the SDW channel saturates whereas the SC channel diverges. We obtain the leading pairing function by decomposing $V^{\rm SC}$ and plot it on the Fermi surface in Fig.\ref{mu116}(a). This is a nodal $s$-wave pairing. Fig.\ref{mu116}(d) shows the SDW interaction at the final stage, showing enhancement at both $\v q_1$ and $\v q_2$, although it is relatively stronger at $\v q_1$. These vectors matches roughly the sign change in the pairing function in Fig.\ref{mu116}(a). This is consistent with the general observation that for singlet pairing, the gap function changes sign across Fermi momenta connected by SDW fluctuations.

\begin{figure}
    \includegraphics[width=8.5cm,trim={1.1cm 8.2cm 1.9cm 8.2cm},clip]{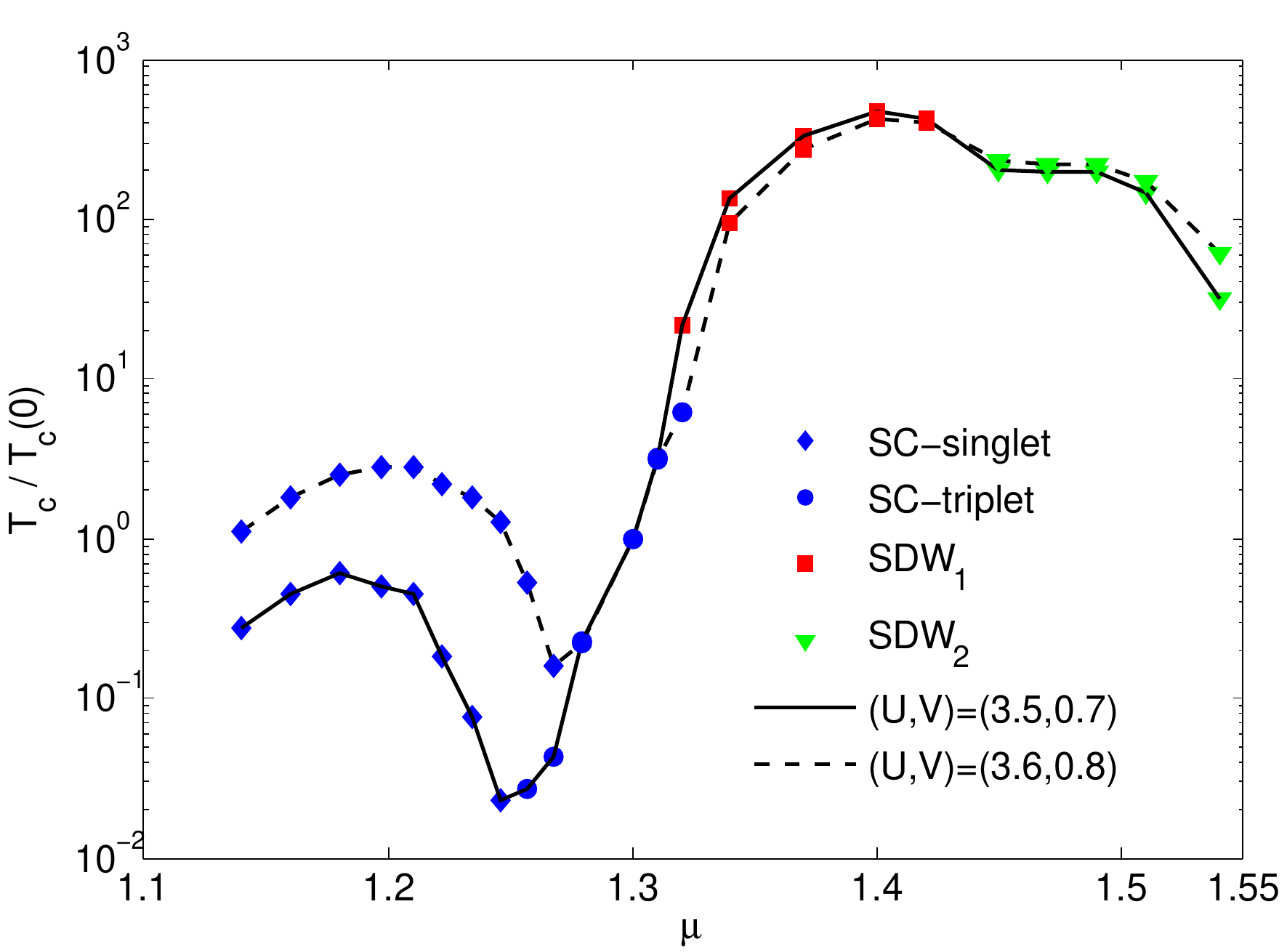}
    \caption{ Normalized transition temperature $T_c$ (symbols) versus $\mu$ from FRG calculations for $(U,V)=(3.5,0.7)$ (solid lines) and $(U,V)=(3.6,0.8)$ (dashed lines). The symbols indicate the respective order that would emerge below $T_c$. Lines are used to guide the eyes.} \label{Tc}
\end{figure}

\begin{figure}
	\includegraphics[width=8.5cm,trim={2.5cm 0cm 2.5cm 0.5cm},clip]{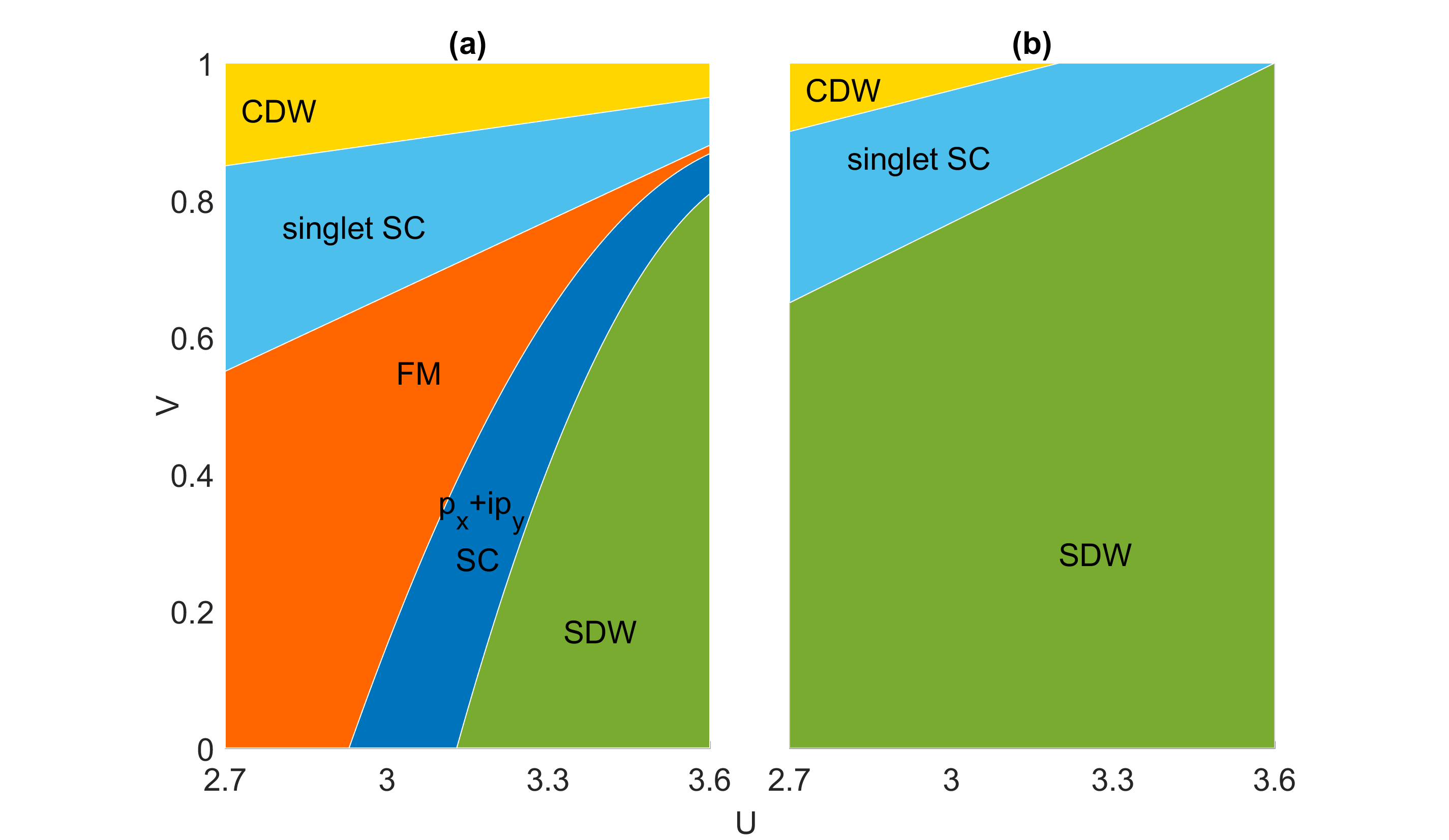}
	\caption{Phase diagram in the $(U,V)$ parameter space. (a) The unstrained case. (b) The case when biaxial strain pushes the system right at the van Hove singularity.} \label{phasediag}
\end{figure}

{\em Phase diagram}: We have performed systematic calculations with respect to various values of $\mu$, for  $(U,V)=(3.5,0.7)$ used above, and for $(U,V)=(3.6,0.8)$ for comparison. In Fig.\ref{Tc} we plot the transition temperature (or the divergence scale in FRG) for the leading orders as a function of $\mu$. We find nodal $s$-wave SC around $\mu=1.2$, a narrow region of $p$-wave SC around $\mu=1.3$, an SDW$_1$ state at small momentum $\v q_1$ near the van Hove level $\mu=1.4$, and finally an SDW$_2$ state at large momentum $\v q_2$ for $\mu>1.42$. The rich phases follow from the drastic change of Fermi surface topology across the van Hove singularity. The two sets of interactions produce qualitatively the same behaviors. The relative transition temperature in the narrow $p$-wave SC region changes by roughly two orders of magnitude. In terms of actual units, we need to take into account the relative change of the hopping integral, which is proportional to the strain. According to experiment a strain level of a few percent is sufficient to drive the system up to (or even beyond) the van Hove level. Therefore even in actual units our result still implies a rapid enhancement of $T_c$ as the strain drives the system toward the van Hove level.

We compare the phase diagrams in the interaction parameter space in Fig.\ref{phasediag} for (a) the unstrained case, and (b) the case when the biaxial strain pushes the system right at the van Hove singularity. (Notice that in the unstrained case the FM phase would yield to $p$-wave SC if the parameters are reduced, but the transition temperature would be too low to be of practical interest.) We find that while there is a sizable regime for $p$-wave SC in case (a), there is no such a phase to our confidence in case (b). However, near the CDW phase boundary (where $V\sim U/4$) there is a regime of singlet SC phase in both cases. 
According to the phase diagram, if the system starts from the singlet SC phase in (a), it may keep in it at the van Hove level in (b) upon biaxial strain. However, if the system starts from the $p$-wave SC phase in (a), it is unlikely to survive at the van Hove level in (b).

Since the van Hove level is of particular interest in experiment, in the Appendix we double check case (b) by reducing the FRG to the more conventional analytical RG (the g-ology RG). We find that SDW (or CDW) dominates for $0<V<U/4$ (or $V>U/4>0$), and singlet pairing occurs only in close proximity to $V=U/4>0$, in agreement with the FRG result. The consistency between different levels of approximation suggests the robustness of the phase diagram (b).

The thermal conductivity measurement in Ref.\cite{kappa} suggests vertical nodes in the gap function. This could be made possible in the $p$-wave SC phase only accidentally, as we discussed above, but is protected by symmetry in the case of singlet $d$-wave SC. However, we find that the gap function in the singlet-SC regime in Fig.\ref{phasediag} is $s$-wave nodal, similarly to that in Fig.\ref{mu116}(a). Although such a nodal structure is also not protected by symmetry, it arises rather naturally because of the coexistence of strong SDW fluctuations at both smaller and larger momenta. Without an accurate estimation of bare interaction parameters we can not rule out the possibility of singlet SC in the real material. But we remark that previous $\mu$SR,\cite{muSR} Kerr effect \cite{Kerr} and phase-sensitive SQUID experiments \cite{liu} strongly support $p$-wave pairing. The Most recent Little-Parks experiment in micro-rings of Sr$_2$RuO$_4$ appears to be consistent with odd-parity pairing. \cite{little}

\section{Summary} \label{SMR}

To conclude, we investigated the possible electronic orders in Sr$_2$RuO$_4$ thin film under planar biaxial strain. We found triplet $p$-wave pairing in a very small region of positive strain, with enhanced transition temperature. However, well before the strain pushes the $\ga$ band up to the van Hove level, or the Lifshitz transition, the system develops small-$\v q$ SDW order, followed by large-$\v q$ SDW order as the Fermi level is pushed even higher. (The large-$\v q$ SDW may interfere with that from the $\al$ and $\bt$ bands ignored here.) We notice that the narrow region of $p$-wave SC may be consistent with the absence of SC in the experiment. \cite{Burganov} This is because the strain depends on the substrate and hence can not be tuned continuously. The smallest experimental strain may have pushed the system outside of the SC region into the SDW regions. However, it is also possible that SC would appear in thin films of better quality to be achieved. We find this is possible if the unstrained system is in a parameter space where singlet SC is realized, which is however inconsistent with previous evidences for $p$-wave SC. Assuming negative strain would lower the Fermi level of the $\ga$ band, we also found nodal $s$-wave SC that could be checked by further experiments.

\acknowledgments{We thank T. M. Rice for discussions. QHW was supported by National Key Research and Development Program of China (under grant No. 2016YFA0300401) and NSFC (under grant No. 11574134). WSW was supported by NSFC (under grant No. 11604168). FCZ was supported by NFSC (under grant No. 11674278) and National Basic Research Program of China (under grant No. 2014CB921203).}

\section{Appendix}

{\em Technical ingredients of SM-FRG}:
Here we present necessary technical details for SM-FRG.  Consider the interaction hamiltonian $H_I=(1/2)c_{1\si}^\dagger c_{2\si'}^\dagger \Ga_{1234} c_{3\si'} c_{4\si}$. Here the numerical index labels momentum/position, and we leave implicit the momentum conservation/translation symmetry. The spin SU(2) symmetry is guaranteed in the above convention for $H_I$. The idea of FRG is to get the one-particle-irreducible interaction vertex $\Ga$ for fermions whose energy/frequency is above a scale $\La$. (Thus $\Ga$ is $\La$-dependent.)
Equivalently, such an effective interaction can be taken as a generalized pseudo-potential for fermions whose energy/frequency is below $\La$. It is useful to  define matrix aliases of the rank-4 `tensor' $\Ga$ via
\eqa \Ga_{1234}=P_{(12)(43)}=C_{(13)(42)}=D_{(14)(32)}.\eea
Here $P$, $C$ and $D$ are matrices of combined indices, reflecting scattering amplitudes for fermion bilinears in the pairing, crossing and direct channels. Starting from the bare interactions at $\La=\infty$, the interaction vertex flows toward decreasing scale $\La$ as,
\eqa \frac{\p \Ga_{1234}}{\p\La} = &&[D\chi^{ph}(D-C)+(D-C)\chi^{ph}D]_{(14)(32)}\nn
&& +[P\chi^{pp}P]_{(12)(43)} - [C\chi^{ph}C]_{(13)(42)},
\label{Eq:dV} \eea
where matrix convolutions are understood within the square brackets, and
\eqa && \chi^{pp}_{(ab)(cd)} = \frac{1}{2\pi}[G_{ac}(\La)G_{bd}(-\La)+(\La\ra -\La)],\nn
&& \chi^{ph}_{(ab)(cd)} = -\frac{1}{2\pi}[G_{ac}(\La)G_{db}(\La)+(\La\ra -\La)],
\label{Eq:def} \eea
where $G$ is the normal state Green's function, and we used a hard-cutoff in the continuous Matsubara frequency. \\

From $\Ga$ (or its aliases $P$, $C$ and $D$), we extract at a given scale $\La$ the effective interactions in the general SC/SDW/CDW channels
\eqa (V^{\rm SC},V^{\rm SDW},V^{\rm CDW}) = (P, -C, 2D-C). \label{eq:channel}\eea
They are matrices describing scattering of fermion bilinears in the respective channels. Since they all originate from $\Ga$, they are overlapped but are naturally treated on equal footing. The effective interactions can be decomposed into eigen modes. For example, in the SC channel (with a zero collective momentum),
\eqa
[V^{\rm SC}]_{(\v k,-\v k)(\v k',-\v k')} = \sum_m f_m(\v k)S_m f_m^{*}(\v k'),
\eea
where $S_m$ is the eigenvalue, and $f_m(\v k)$ is the eigenfunction, which can be expanded in terms of lattice harmonics, such as $e^{i\v k\cdot \v r}$ where $\v r$ is the distance between the fermions within a fermion bilinear. We look for the most negative eigenvalue, say $S=\min[S_m]$, with an associated eigenfunction $f(\v k)$. If $S$ diverges at a scale $\La_c$, it signals the instability of the normal state toward a SC state, with a pairing function described by $f(\v k)$. Similar analysis can be performed in the CDW/SDW channels, with the only exception that in general the collective momentum $\v q$ in such channels is nonzero. Since $\v q$ is a good quantum number in the respective channels, one performs the mode decomposition at each $\v q$. There are multiple modes at each $\v q$, but we are interested in the globally leading mode among all $\v q$. In this way one determines both the ordering vector $\v Q$ and the structure of the order parameter by the leading eigenfunction. Finally, the instability channel is determined by comparing the leading eigenvalues in the CDW/SDW/SC channels.\\

In principle, the above procedure is able to capture the most general candidate order parameters. In practice, however, it is impossible to keep all elements of the `tensor' $\Ga$ for computation. Fortunately, the order parameters are always local or short-ranged. This is notwithstanding the possible long-range correlations between the order parameters. For example, the s-wave pairing in the BCS theory is local, since the gap function is a constant in momentum space. The order parameter in usual Landau theories are assumed to be local. The d-wave pairing is nonlocal but short-ranged. The usual CDW/SDW orders are ordering of site-local charges/spins. The valence-bond order is on-bond but short-ranged. In fact, if the order parameter is very nonlocal, it is not likely to be stable. The idea is, if it is not an instability at the tree level, it has to be induced by the overlapping channel. But if the induced order parameter is very nonlocal, it must be true that the donor channel has already developed long-range fluctuations and is ready to order first. These considerations suggest that most elements of the `tensor' $\Ga$ are irrelevant in the RG sense and can be truncated. \Eq{Eq:dV} suggests how this can be done. For fermions, all 4-point interactions are marginal in the RG sense, and the only way a marginal operator could become relevant is through coherent and repeated scattering in a particular channel, in the form of convolution in \Eq{Eq:dV}. Therefore, it is sufficient to truncate internal spatial range within the fermion bilinear, e.g., between 1 and 2, and between 3 and 4, in  $P_{(12)(34)}$. This means that the form factors are expanded in a truncated set of lattice harmonics. The setback distance between the two groups is however unlimited (thus thermodynamical limit is not spoiled). Similar considerations apply to $C$ and $D$. Eventually the same type of truncations can be applied in the effective interactions $V^{\rm CDW/SDW/SC}$. Such truncations keep the potentially singular contributions in all channels and their overlaps, underlying the key idea of the SM-FRG.~\cite{WSW1, YYX,Wang2014} The merits of SM-FRG are: 1) It guarantees hermiticity of the truncated interactions; 2) It is asymptotically exact if the truncation range is enlarged; 3) It respects all underlying symmetries, and in particular it respects momentum conservation exactly. 4) In systems with multi-orbitals or complex unitcell, it is important to keep the momentum dependence of the Bloch states, both radial and tangential to the Fermi surface. This is guaranteed in SM-FRG since it works with Green's functions in the orbital basis. These are important but may be difficult to implement in the more conventional patch-FRG applied in the literature.~\cite{Honerkamp2001, Metzner2012, Platt2013}\\

To check the convergence of the real-space truncation for fermion bilinears discussed above, we define $L_c$ as the maximal distance between the two fermions within a fermion bilinear. We take a sufficiently large $L_c$ such that the results are not sensitive to a further increase of $L_c$. In the main text, we used $L_c$ up to the forth-neighbor bond.\\

{\em G-ology at the van Hove level}: Since the van Hove level is of particular interest in the experiments, here we also present analytical RG (or g-ology) at this level, focusing on momenta near the saddle points. It can be taken as a tremendous  approximation or reduction of FRG: all momentum dependence in the interaction vertices are projected on the saddle points. It applies best when the divergence scale is sufficiently low to warrant the approximation around the saddle points. Since there are but two independent saddle momenta, which we label as `1' and `2', there are only four independent vertices: 
\eqa (g_0, g_u, g_b, g_f) = (\Ga_{1111}, \Ga_{1122}, \Ga_{1212}, \Ga_{1221}),\eea
representing intra-saddle, umklapp, back-scattering and forward-scattering interactions, respectively. The RG equations can be written as, within logarithmic accuracy and up to an umimportant global factor (which can be absorbed by rescaling the interactions), 
\eqa && d g_0/dt = -(g_0^2+g_u^2)t + g_0^2+g_b^2-2 g_f^2+2 g_b g_f, \nn 
&&d g_u/dt = -2g_u g_0 t + 4 \al g_u g_f - 2\al g_u g_b, \nn 
&&dg_b/dt = -2\al g_b^2 + 2 g_0 g_b, \nn 
&&dg_f/dt = -\al g_b^2 +\al g_u^2 + 2g_0 g_b - 2g_0 g_f,\eea
where $t=\ln(E_0/\La)$ is the RG time ($E_0\sim 1$ is a starting energy scale), and $\al$ is a factor reflecting the degree of nesting of the Fermi surface near the saddle points. For our case the nesting is weak (and this is why ferromagnetic correlations are favored) and we take $\al \sim 0.5$. The initial values are given by
\eqa (g_0, g_u, g_b, g_f) = (U+4V, U-4V, U-4V, U+4V).\eea
For $0<V<U/4$,  as the RG flows to strong coupling, we find the singlet pairing interaction $V_{SC}=g_0\pm g_u >0$ for $s$ or $d$-wave symmetry. Consistently, the ferromagnetic interaction $V_{FM}=-(g_0+g_b) <0$. For $V>U/4>0$ we end up with CDW. Exactly at $V=U/4$ the $s$-wave pairing, $d$-wave pairing and the Pomeranchuk mode (a $d$-wave CDW at zero momentum) are leading and degenerate. (This high degree of degeneracy may be an artefact because of the over simplification in g-ology.) For $V$ immediately below (above) $U/4$, we find $d$-wave ($s$-wave) pairing to be the leading instability. Finally, triplet pairing is forbidden since the saddle point momenta are time-reversal invariant. The overall feature is consistent with Fig.\ref{phasediag}(b) in the main text. 

Interestingly, we may also apply the g-ology to the van Hove level caused by the uniaxial strain. In this case there is only one saddle momentum. If we project all momentum dependence in the interaction vertices on this saddle point, we end up with just one independent vertex, $g_0=\Ga_{1111}$. The flow equation can be solved trivially. The result is any repulsive $U$ and $V$ are screened. This appears to be too afar from both experiment and FRG. We can do better by including two more Fermi momenta, say $L$ and $R$, orthogonal to the saddle momentum. In this setting we have seven independent vertices, but we find for weak $0<V\ll U/4$, all vertices converges during RG flow, except that $\Ga_{LLLL}$ (or $\Ga_{RRRR}$) diverges linearly in $t$ (or logarithmically in $\La$). On one hand this means all interaction vertices are marginal. On the other hand, the pairing interaction (determined by $\Ga_{1111, LRLR, LRRL}$) is finite and hence there is no SC phase, while the FM interaction (dominated by $-\Ga_{LLLL, RRRR}$) is attractive and diverges, although only logarithmically. Although this analysis predicts ferromagnetism as the leading interaction, the logarithmic divergence does not imply an emerging order. We think this failure is an artefact of the over-simplified g-ology. In this case, the FRG as applied in Ref.\cite{YCL} and in the main text becomes indispensable.

\end{document}